# Non-Volatile Kernel Root kit Detection and Prevention in Cloud Computing

R.Geetha Ramani , S Suresh Kumar

*Abstract: The field of web has turned into a basic part in everyday life. Security in the web has dependably been a significant issue. Malware is utilized to rupture into the objective framework. There are various kinds of malwares, for example, infection, worms, rootkits, trojan pony, ransomware, etc. Each malware has its own way to deal with influence the objective framework in various ways, in this manner making hurt the framework. The rootkit may be in some arbitrary records, which when opened can change or erase the substance or information in the objective framework. Likewise, by opening the rootkit contaminated record may debase the framework execution. Hence, in this paper, a Kernel Rootkit Detection and Prevention (KRDP) framework is proposed an avert the records. The avoidance system in this paper utilizes a calculation to forestall the opening of the rootkit influenced record as portrayed. By and large, the framework comprises of a free antivirus programming which is restricted to certain functionalities. The proposed model beats the functionalities by utilizing a calculation, in this way identifying the rootkits first and afterward cautioning the client to react to the rootkit tainted record. In this way, keeping the client from opening the rootkit contaminated record. Inevitably, in the wake of expelling the tainted document from the framework will give an improvement in the general framework execution.*

*Index Terms: Cloud; file; malware; port; rootkits; process; prevention;*

## I. INTRODUCTION

Spread enrolling cloud give undeniable relationship, for instance, SaaS, PaaS, and IaaS, to clients over the web. Windows precedent (AWS) are ruined by the aggressors with rootkit over the electronic world. Rootkit [1] settlement expect an essential occupation in various malware [7][8] tests to cover the exercises of the aggressors, for instance, remote access, keylogging, records, and procedure. This issue had been raised for a long time. One risky strike in a rootkit, which kind of malware, can engage also advantaged access to a PC. Meanwhile, it could cover the closeness of itself and obvious toxic technique by revealing some improvement and modification to the structure. Thus, ID of rootkit isn't immediate as that of various malware. The bit rootkits change the decided resources of bit, which deals the reasonableness of the OS section. Anyway various investigates go for seeing section rootkits, perceiving

R.Geetha Ramani, Professor, Department of Information Science and Technology, College of Engineering, Anna University, Chennai, India,

S Suresh Kumar, Associate Professor, Department of Computer Science and Engineering, Rajalakshmi Engineering College, Chennai

rootkits in cloud condition ought to be clear to cloud clients, without adjustment to applications and OS.

Specialists have made contraptions, for instance, GMER [2], RootRepeal, VBA32 arkit, Rootkit Unhooker, Dr.Webcurelt, IceSword [2] to see hidden [5] contradict. They perceive covering conduct by looking article got from an unordinary state see with those removed from low-level framework data upon clients request or dependably. To hoard low-level information, an extensive bit of these instruments rely on unequivocal section data structures that are used by the working framework, fundamentally to taking a gander at and bookkeeping purposes. These data structures are, regardless, every now and then exceedingly framework subordinate, and may change across over different framework outlines or even fix levels. Additionally, if rootkit makers later find a few courses of action concerning such data structures, they can change their rootkits to control these data structures and in this way avoid revelation. Along these lines, these affirmation mechanical congregations may wrap up vain when rootkits advantage novel covering methodology.

The epic covering procedures is a cross view based territory structure which vanquishes a dash of the detainment of existing approachs. This framework goes for seeing records masked by rootkits. It works by never-endingly keeps up a snappy diagram of records outside the checked structure, that mirrors the bona fide running framework. Part Rootkit Detection and Prevention (KRDP) joins three sections: Pattern Checker, Byte Generator and the Byte Analyser

## II RELATED WORK

Circulated processing is amazing as "Pay Per Use" and "On-Demand" figuring[3]. It is a kind of Internet-based selecting which offers pooled organizing resources and data, to PCs and other figuring contraptions on intrigue. Cloud is standard model for allowing on-demand, determined access to a typical get-together of configurable enrolling resources, for instance, information aggregating, association, server, application and systems, which can be speedily provisioned and freed with minor running effort. Scattered preparing suppliers supply their "influence" in 3 explicit models, Infrastructure as an association (IaaS), Platform as a Service(PaaS), and Software pack library as a Service(SaaS). Appropriated enrolling can be made into four sorts. They are open, private, framework and mutt cloud .This examination basically revolves around scattered enlisting headway, and therefore requires a formal hugeness of passed on figuring. The display of using a game-plan of remote servers encouraged on the Internet to perform various exercises, for instance, putting away, the board,
and treatment of information,

instead of on a region server or a PC.

Rootkit [1]: This one is connected with the "criminal stowing unendingly in the additional room, keeping a tight grasp on take from you while you are not in home".

It is the hardest of all Malware to see and in this manner to clear, unique prodigies propose inside and out cleaning your hard drive and reinstalling everything beginning with no outside assistance. It is depended upon to permit other information gathering Malware in, to get the character information from your PC without you understanding anything is going on. Such Malware are being made by those endeavoring to get to the PC for cash related extension. As such, Rootkit zone in spread preparing association recognize a basic development. There are a few past examinations related to this examination dealing with all of scattered preparing and structure and what's more prominent certification frameworks.

The current rootkit revelations, (signature, lead, cross view, tolerability, and equipment) delineate these methodologies. The engraving based exposure technique is most dependably used procedure to see rootkit. Right when antivirus makers gain a touch of malware, it see a "signature" that is extraordinary to the byte case of the malware and distinguish those models in a check database. Territory programming bringing the point of reference from the framework and those models are showed up diversely in connection to the engravings in the database. If there is a match, the database sees the malware and it won't separate any malware that does not support an engraving in the database. The shortcoming of engraving based affirmation, is that it works only for known malware and an attacker can disable the firewall before demonstrating the rootkit. The upside of engraving based rootkit is that it works only for known rootkits that stow away in the memory. The other unmistakable confirmation methodology don't see rootkits in memory and additionally signature-based strategies .Behavior-based rootkit zone routinely sees new rootkits that don't yet have known engravings. It picks the direct of the given framework and looks for the refinements. Those abilities can be run of the mill for malware on a structure.

In cross view rootkit affirmation, the two extraordinary points of view of the system are showed up differently in association with find contrasts. This framework gets an unusual state point of view of the structure from a district that is weak to control by malware [4]. The sporadic state see won't report whatever the rootkit stows away. This procedure sees the outside view as the true blue point of view of the structure since it demonstrates the veritable view on the framework instead of expecting the framework reports is liberal. Clean booting framework is the best way to deal with oversee find stealth malware and to keep the malware from covering itself. It can't cover records or method from an outside source gave that the OS isn't running, the malware isn't running. From boot menu, the perfect boot technique boots the framework into DOS mode for Windows9x system and also looks record structure. For Windows ME, clean booting requires outside instruments to boot the structure into DOS mode.

An elective course of action has been proposed by Jonathan Grimm et.al[3], which looks to ordinarily see and debilitate rootkit malware by reestablishing standard framework control streams. The structure utilized virtual machine care (VMI), which permits a remarkable VM to see and control the physical memory of different VMs with the guide of the hypervisor.

Conventionality based affirmation looks confided in model, got when the framework was impeccable, to the present viewpoint of memory or the record structure. Any refinements can show a rootkit's quality on the machine. Apparatus based rootkit divulgence began from the likelihood that outside equipment would not battle with the rootkit for resources, for example, programming based distinctive evidence does. Like outer programming, outside equipment similarly screens structure movement at a lower level. The advantage of equipment rootkit territory is, the rootkits can't modify the rigging since it utilizes an outside OS. This paper displays a cross-see clean boot based disclosure technique which gags a pinch of the suffering methodologies of rootkit. The proposed strategy goes for seeing records, covered up by rootkits.

### III. SYSTEM DESIGN

Presence of hidden files is detected by using the proposed system KRDP design as shown in Figure 1.

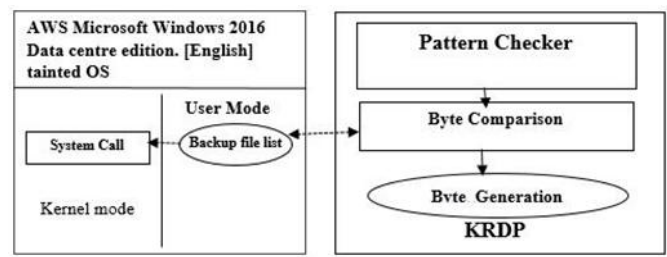

**Figure 1 System Overview of Kernel Rootkit Detection and Prevention (KRDP)**

The structure design includes three major modules. Each module has it's own one of a kind task to perform given by the customer. It incorporates two records, the principal report and the supported up record (before impacted by rootkit). The general system incorporates sending two archives through the model checker and separating it to see whether there's any clues of rootkit concealed inside the record. In case there are any clues of rootkit in the report, by then the records are sent to byte age. In byte age, the reports in the wake of checking it gets evacuated and byte estimations of records are resolved. The byte estimations of the record are in numerical association. The connection is done with a conventional record and the bolstered up report. After the examination is done, in case the both the byte regards for both the records are same, by then there are possible results that the archive may be impacted by the rootkit. By then the records are moved to byte relationship. The byte examination is used to take a gander at the byte estimations of each archive by differentiating each byte estimation of the record. If the byte estimations of the records are interesting, by then it is deduced that the archives are affected by rootkit. The proposed system is used to guarantee that the rootkit is perceived.
The customer is frightened,
along these lines the customer

can remove the rootkit record from the structure. The animated synopsis can be gotten from the userland application occasion. This is used to build up the rundown synopsis of polluted OS and it ought to be made as sporadic state as could reasonably be ordinary, to absolutely get the view seen by customers.

*A. Pattern checker*

The model checker module contains two records, the archive in the principal coordinator and the report in the maintained up envelope. The bolstered up envelope contains archives that are maintained up before impacted by contamination. The report is checked for the rootkit configuration by registering the SHA-256[6] estimations of the record and checking whether it is comparable or not and there by finding the rootkit plan in it. A SHA hash a motivation for each record can be resolved and if the archive is modified, by then its SHA-256 characteristics changes. Thusly, if the primary record is struck by rootkit the hash estimations of the reports will change. Secure Hash Algorithms (SHA) are a class of cryptographic limits that is expected to keep data confirmed. In Secure Hash Algorithms the data is changed into work regards by playing out explicit limits that incorporates bitwise errands, estimated increments and weight limits. There are three extraordinary sorts of SHA works so far explicitly, SHA-1, SHA-2 and SHA-3. Every limit conveys a yield as a hash regard that fluctuates in number of bits. Here SHA256 computation is used. It makes an exceptional 256 piece (32 byte) signature for a substance or a number. The hash estimations of the principal archive and the maintained up record is resolved. Each rootkit has a SHA regard which addresses the model or imprint. The archive is checked for the rootkit configuration by finding out the SHA256 regard for the first and the maintained up record and after that taking a gander at the hash estimations of the reports. In case the characteristics differentiate, by then the contamination configuration matches with the record. In this paper we have considered a model report and decided hash a motivating force for that record. Underneath given the pseudocode that is used for checking rootkit plan in the record. Figure 3 Pattern checker yield appeared.

**Pseudo Code**

```
function PatternChecker()
{
Isequal = true;
Set hex1 to checksum()
Set hex2 to checksum()
for( i:= 1 to length(hex1)
        do
{
If (Hex1[i] Compare Hex2[i])
Isequeal to false;
}
}
```

In the above pseudo code, the checksum() is the method that calculates SHA256 [6] value for a file.

In the principle work the an item md is made which is the case of class MessageDigest that gives the usefulness of a message digest calculation, for example, MD5 or SHA. The getInstance() strategy is utilized determine the kind of calculation to be utilized. At that point the checksum technique is called by passing the way of the document and the message condensation object md. Inside the checksum technique, the document hashing is done and the outcome is put away in string developer as a string. The outcome is returned as a string to the primary capacity. At that point both the hash esteems in the string are checked whether they are equivalent or not by looking at character by character. On the off chance that they are equivalent, the framework is protected and there isn't infection design in the record. In the event that the qualities vary, at that point it demonstrates that the infection example matches with the document.

*B. Byte Generation and Byte Comparison*

If the rootkit configuration matches with the archive, by then the records are sent to byte age where the records are changed over to bytes. Each record will have a byte regard. If the substance of the archive changes, by then the byte estimations of the record in like manner changes. In case the record is attacked by rootkit, by then the substance of the archives are in like manner changed and in this way the byte regards are also changed. Along these lines, the byte estimations of the archive are used as a seat engraving to check if the record is a rootkit affected report. Underneath given the code for byte age and connection. Figure 4 indicated byte examination yield.

**Pseudo Code**

```
function BgBc()
{
Isequal = true;
Create cleanfile byte array for original file;
Create taintedfile byte array for tainted file;
i-0;
j=0;
 while(I<cleanfile.length()&&j< taintedfile.length())
        {
          If ( cleanfile != taintedfile)
            do
             {
              isequal=false;

              i++;
              j++;
         }
      If isequal==true;
        printf("The bytes value of the files have unchanged");
else
printf("The bytes value of the files have changed! The file is affected by Rootkit");
}
function readbytefromfile(str filepath)
{
try
{
```

```
        File file= new File(path);
//Create file pointer Bytearray = new Byte[(int) file.length()]
//Convert byte  a array
        Filepointerpointout to inputstream
            }
        catch(exception e)
            { }
    return bytearray;
    }
```

In this pseudo code, two byte clusters are made which speaks to the records for which the rootkit example is checked. The byte clusters are passed to readbytesfromfile technique. This technique makes a byte cluster of same length and changes over the record to byte and returns the byte exhibit to the calling capacity. At that point in the primary capacity, both the byte clusters are contrasted with check whether they are equivalent or not. On the off chance that they are extraordinary, at that point we can infer that the document has been influenced by rootkit.

## IV. IMPLEMENTATION

AWS is introduced by signing into the AWS Management Console and setting up root account. At that point in the Amazon EC2 Dashboard, pick "Dispatch Instance" to make and design your virtual machine. In this wizard, you have the choice to arrange your occurrence highlights. It might take a couple of minutes to introduce your case. Associate the case. We can interface by means of SSH, PUTTY and RDP. We can end it by choosing the EC2 case, pick "Activities", select "Example State", and "End".

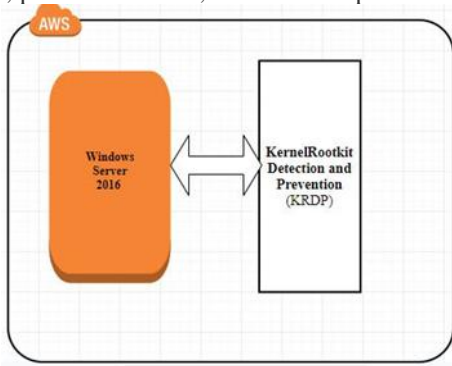

**Figure 2 Implementation of Kernel Rootkit Detection and Prevention**

The execution technique is showed up in the layout in Figure 2. In this examination, KRDP is endeavored on rootkit tests that hold running under an AWS cloud condition. First run the KRDP on a (first occasion ruined OS) Windows Server 2016 with an Intel Xeon Family 2.5 GHz CPU and 1GB RAM. The system is a standard establishment with organization pack one. For this circumstance, the windows Server 2016 working system is used. To incorporate the working structure, open AWS are and open the downloaded windows Server 2016 iso record and starting advances is done to finish the setup. The AWS is gotten to, windows Server 2016 working structure is opened. Before instating the working system, the common report decision is enabled by going to properties and picking enable shared record get to elective. Further, the record path zone to get to the specific archives and lists is picked. This will empower the customer to get to the working structure records inside the woring system through AWS.

The windows Server 2016 working system is presented in AWS. Backing is opened and any java record foundation invigorates that is required are downloaded and presented. Before the guest working structure is defiled with rootkit, a fortification of all the piece records and a clone of windows Server 2016 guest working system is taken. Test rootkit are used to debase the working system. This where the "KRDP "obstructs the methodology. In this system the "KRDP "involves three modules. Model Checker, Byte Generation and Byte Comparison. The figuring is accessible inside the "KRDP" module. Rootkit could corrupt any archives, for this circumstance, the bit records are affected. The reports experience the Pattern Checker for distinguishing rootkit marks. The model checker is the starter test. Further, the Byte Generation is done to deliver the byte size of each record and address in bytes. Further, Byte Comparison is done to compere the support archives with defiled records. If the record measure is changed, by then, the customer gets proposal about the rootkit defiled report. Along these lines, the "KRDP" computation shields the customer from opening a rootkit defiled record by forewarning the customer.

| FILE NAME | TYPE OF FILE IT AFFECTS | SIZE OF THE FILE |
|---|---|---|
| Virus.Boot.IsraeliBoot.a | Kernel.dll | 512 bytes |
| Virus.Win32.Enerlam.c | Kernel32.dll | 6KB |
| Virus.Boot.Gwar | Tasthost.exe | 512 bytes |
| Virus.Win32.Orez.6291 | User32.dll | 12.0 KB |
| Virus.Win32.Orez.6287 | Ucrtbase.dll | 12.0 KB |
| Virus.Multi.b | System32.dll | 68 KB |
| Virus.DOS.Vienna.Violator.699 | Scvhost.exe | 1.4 KB |
| Virus.BAT.Batman.b | Auto.bat | 256 bytes |
| Virus.Win32.Hawey | Scvhost.exe | 5.5 KB |

**Table 1 Sample Rootkit**

## V. PERFORMANCE

PC Performance is characterized by the quantity of undertakings achieved by the PC framework. Execution can be depicted as far as proficiency, speed, and precision. The capacity to play out the given assignments consistently and finishing in a shorter time makes the framework to be all the more dominant in execution. In the event that a framework ought to be higher in it's exhibition, at that point, it ought to have higher transfer speed, less information transmission time, quicker information pressure and decompression, lower use of processing assets and capacity, higher throughput and shorter holding up time. For this situation, the AWS is utilized for testing and the visitor working framework utilized is Windows server 2016. Before the infection contaminated the visitor working framework, a reinforcement of the considerable number of records were taken. The framework execution speaking to the CPU usage, Processes, Threads, and Handles preview was taken before the framework was contaminated with infection. Further, after the framework was contaminated with infection, the records and framework execution after disease was contrasted and the unaffected documents and reinforcement taken before. Whenever looked at, there was an expansion in the CPU usage. Prior it was 17%, yet when contrasted and the present CPU usage, there was an expansion of 8%, along these lines making it as 25%. The quantity of procedures before influenced were higher than the quantity of procedures after influenced, however the quantity of strings expanded. Before the framework got tainted, there were 652 strings, yet after the framework got contaminated, the quantity of strings expanded by 35, making it 687. The quantity of handles before influenced were 16,720 and after the framework was contaminated, there was an expansion of 53 handles, making it 16,773 handles. Memory use before tainting was 0.9 GB out of 2.0 GB. In the wake of tainting the framework, the Memory usage is 0.7 GB out of 2.0 GB. The usage of Memory may be shockingly low after disease, however that is expected to the rootkit expelled. Prior, the rootkit was stowing away, in this way it had more utilization of Memory. After the rootkit began to contaminate, the Memory usage was less when contrasted and the past framework execution. The proposed calculation had the option to distinguish the rootkit record. Further, the client is told by rootkit movement. By utilizing KRDP, the rootkit document is informed to the client, which after expulsion by the client gives an expanded framework execution. Along these lines, the exhibition is reestablished to its ordinary state.

## VI. EXPERIMENTAL RESULTS

The algorithm is tested for its correctness, by testing it with some sample rootkit that are collected from the offensive computing website [7] [8]. The algorithm worked well for all the rootkit samples and was able to detect all the virus samples that are listed above in the Table 1.

**Table 2 Comparison of size of Sample file before and after affected**

| NAME OF THE SAMPLE FILE | SIZE OF THE FILE BEFORE AFFECTED | SIZE OF THE FILE AFTER AFFECTED |
|---|---|---|
| KEYBOARD.SYS | 9 KB | 42KB |

The algorithm was executed in command prompt and the output was true for pattern checker and as a result of this, the byte values of the file have changed. Hence we concluded that the file was affected by rootkit. Table 2 Comparison of size of Sample file before and after affected shown. The output is shown as follows.

**Figure 3 Pattern checker output**

**Figure 4 Byte Comparison output**

## VII. CONCLUSION AND FUTURE WORK

Rootkit discovery is a topical subject area. It has been created to conquer the issues related with customary mark based infection discovery. In this task, we concentrated on discovery of an infection influenced document and anticipation of the spread of rootkit record by keeping the document from opening. For the most part, a rootkit that influences records spread when it is opened. So its task is ceased by structuring an avoidance model that utilizations two dimensions of checking to affirm the nearness of rootkit in a document. This venture has focused more on rootkit location in light of the fact that once it is recognized, the spread of the rootkit can be halted by not opening it. So this is in reality first dimension of counteractive action. The test results is right and appropriate to identify the rootkit. In future, we will attempt to counteract the procedure that is executed by the rootkit with the goal that it very well may be killed and attempt to recoup the document and will actualize this model in a cloud domain.

**REFERENCES**

1. R.Geetharamani, S.Sureshkumar, ShomonaGracia Jacob, "Rootkit (Malicious Code affecting Kernel) Prediction through Data Mining Methods and Techniques" , *IEEE-ICCIC-2013*.
2. LeianLin , Zuanxing Yin, Yuli Shen Haitao Lin, "Research and Design of Rootkit Detection Method" , Elsevier, ICMPBE, 2012.
3. Safaa Salam Hatem, Dr.Maged H, wafy, Dr.Mahamoud M. El-Khouly. "Malware Detection in Cloud Computing", .IJACSA, Vol.5, No. 4, 2014.
4. Gerard wagener,Radu State, Alexandre Dulauno, "Malware Behaviour analysis", Springer.
5. Sebastian Eresheim. "The Evolution of process Hiding Techniques in Malware-Current Thread and Possible Countermeasures", Information Processing Society of Japan Sep 2017.
6. Diana Toma, Dominique Borrione, "SHA Formalization", TIMA Laboratory, Grenoble, France.
7. http://www.offensivecomputing.net/
8. https://www.virustotal.com/#/home/upload
9. K. Vijayakumar and C. Arun, "Continuous Security Assessment of Applications in Cloud Environment", International Journal of Control Theory and Applications, ISSN: 0974-5645 volume No. 9(36), Sep 2016, Page No. 533-541.
10. R.Joseph Manoj, M.D.Anto Praveena, K.Vijayakumar, "An ACO–ANN based feature selection algorithm for big data", Cluster Computing The Journal of Networks, Software Tools and Applications, ISSN:

1386-7857 (Print), 1573-7543 (Online) DOI: 10.1007/s10586-018-2550-z, 2018.
11. K. Vijayakumar and V. Govindaraj, "An Efficient Communication Technique for Extrication and Cloning of packets on cloud", International Journal of Applied Engineering Research, ISSN 0973-4562 Vol. 10 No.66 May 2015

## AUTHORS PROFILE

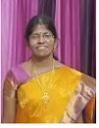
**R. Geetha Ramani** is Professor in the Department of Information Science and Technology at College of Engineering, Anna University, Guindy, Chennai, India. She has more than 20 years of teaching and research experience. Her areas of specialization include Data mining, Bio-informatics, Evolutionary Algorithms and Network Security cloud computing. She has over 100 publications in International Conferences and Journals. She has served as a Member in the Board of Studies of Pondicherry Central University. She is presently member in the Editorial Board of various reputed International Journals

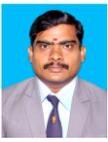
**S. Suresh Kumar** is an Associate Professor in the Department of Computer Science and Engineering, Rajalakshmi Engineering College , Thandalam, Chennai, India. He obtained his Bachelor of Engineering in Computer Science and Engineering from University of Madras, Chennai and Master of Engineering in Computer Science and Engineering from Anna University Chennai .His current research is focused on cloud computing Security.